\documentclass[10pt,twoside,twocolumn,english,aps,manuscript,superscriptaddress,prbt,aps, preprint,manuscript,showpacs]{revtex4}
\usepackage{lmodern}
\usepackage[T1]{fontenc}
\usepackage[latin9]{inputenc}
\usepackage{color}
\usepackage{babel}
\usepackage{array}
\usepackage{float}
\usepackage{amstext}
\usepackage{graphicx}
\usepackage{esint}
\PassOptionsToPackage{normalem}{ulem}
\usepackage{ulem}
\usepackage[unicode=true,pdfusetitle,
 bookmarks=true,bookmarksnumbered=false,bookmarksopen=false,
 breaklinks=false,pdfborder={0 0 1},backref=false,colorlinks=false]
 {hyperref}
\usepackage{breakurl}

\makeatletter

\providecommand{\tabularnewline}{\\}

\@ifundefined{textcolor}{}
{%
 \definecolor{BLACK}{gray}{0}
 \definecolor{WHITE}{gray}{1}
 \definecolor{RED}{rgb}{1,0,0}
 \definecolor{GREEN}{rgb}{0,1,0}
 \definecolor{BLUE}{rgb}{0,0,1}
 \definecolor{CYAN}{cmyk}{1,0,0,0}
 \definecolor{MAGENTA}{cmyk}{0,1,0,0}
 \definecolor{YELLOW}{cmyk}{0,0,1,0}
}
\newenvironment{lyxlist}[1]
{\begin{list}{}
{\settowidth{\labelwidth}{#1}
 \setlength{\leftmargin}{\labelwidth}
 \addtolength{\leftmargin}{\labelsep}
 }}
{\end{list}}

\@ifundefined{date}{}{\date{}}
\makeatother

\begin{document}

\title{Sc$_{2}$Ga$_{2}$CuO$_{7}$: A possible Quantum spin liquid near
the percolation threshold}

\author{R. Kumar }

\affiliation{Department of Physics, Indian Institute of Technology Bombay, Powai,
Mumbai 400076, India}

\author{P. Khuntia}

\affiliation{The Ames Laboratory, US Department of Energy, Ames, IA 50011, USA}

\affiliation{Max-Planck Institute for Chemical Physics of Solids, 01187 Dresden,
Germany}

\author{D. Sheptyakov}

\affiliation{Laboratory for Neutron Scattering and Imaging, Paul Scherrer Institut,
5232 Villigen PSI, Switzerland}

\author{P.G. Freeman}

\affiliation{Laboratory for Quantum Magnetism (LQM), Ecole Polytechnique Federale
de Lausanne (EPFL), CH 1015, Switzerland}

\affiliation{Jeremiah Horrocks Institute for Mathematics, Physics and Astronomy,
University of Central Lancashire, Preston, PR1 2HE U.K.}

\author{H.M. Rønnow}

\affiliation{Laboratory for Quantum Magnetism (LQM), Ecole Polytechnique Federale
de Lausanne (EPFL), CH 1015, Switzerland}

\affiliation{Neutron Science Laboratory, Institute for Solid State Physics (ISSP),
University of Tokyo, Kashiwa, Japan}

\author{B. Koteswararao}

\affiliation{School of Physics, University of Hyderabad, Hyderabad 500046, India}

\affiliation{Department of Physics, Indian Institute of Technology Bombay, Powai,
Mumbai 400076, India}

\author{M. Baenitz }

\affiliation{Max-Planck Institute for Chemical Physics of Solids, 01187 Dresden,
Germany}

\author{M. Jeong}

\affiliation{Laboratory for Quantum Magnetism (LQM), Ecole Polytechnique Federale
de Lausanne (EPFL), CH 1015, Switzerland}

\author{A.V. Mahajan}

\email{mahajan@phy.iitb.ac.in}

\affiliation{Department of Physics, Indian Institute of Technology Bombay, Powai,
Mumbai 400076, India}

\date{{\today}}
\begin{abstract}
Sc$_{2}$Ga$_{2}$CuO$_{7}$ (SGCO) crystallizes in a hexagonal structure
(space group: $P$$6$$_{3}/mmc$),\textcolor{black}{{} which can be
seen as an alternating stacking of single and double triangular layers.
Combining neutron, x-ray, and resonant x-ray diffraction we establish
that the single triangular layers are mainly populated by non-magnetic
Ga$^{3+}$ ions (85\% Ga and 15\% Cu), while the bi-layers have comparable
population of Cu$^{2+}$ and Ga$^{3+}$} ions \textcolor{black}{(43\%
Cu and 57\% Ga)}. Our susceptibility measurements in the temperature
range 1.8 - 400 K give no indication of any spin-freezing or magnetic
long-range order (LRO). We infer an effective paramagnetic moment
$\mu_{eff}=1.79\pm0.09$ $\mu_{B}$ and a Curie-Weiss temperature
$\theta_{CW}$ of about $-44$ K, suggesting antiferromagnetic interactions
between the Cu$^{2+}$($S=1/2$) ions. Low-temperature neutron powder
diffraction data showed no evidence for LRO down to 1.5 K. In our
specific heat data as well, no anomalies were found down to 0.35 K,
in the field range 0-140 kOe.\textcolor{black}{{} The magnetic specific
heat, $C_{m}$, exhibits a broad maximum at around 2.5 K followed
by a nearly power law $C$$_{m}$$\propto$ $T$$^{\alpha}$ behavior
at lower temperatures, with $\alpha$ increasing from }0.3\textcolor{black}{{}
to} 1.9\textcolor{black}{{} as a function of field for fields upto 90
kOe and then remaining at 1.9 for fields upto 140 kOe. Our results
point to a disordered ground state in SGCO. }
\end{abstract}


\maketitle
\textit{Introduction: }\textcolor{black}{Geometrically frustrated
spin systems are full of surprises and continue to draw attention
because of their tendency to host novel ground states \cite{1 Balents}.
In particular, triangular lattice based geometrically frustrated Mott
insulators have gained wide interest because their ground states were
envisaged to be based on a Resonating Valence Bond (RVB) picture \cite{2 Anderson}.}
Anderson's RVB proposal \cite{2 Anderson} pertained to the two dimensional
(2D) edge-shared triangular lattice which, however, has a 120$^{\circ}$
ordered ground state \cite{3 NEEL Support,4 NEEL Support,5 NEEL Support,6 NEEL Support}.
This, in particular, triggered the search for the RVB state (also
called the quantum spin liquid QSL state \cite{1 Balents}) in geometries
such as triangular (anisotropic), Kagomé (2D), hyperkagomé (three
dimensional 3D), and pyrochlore (3D). Finally, a QSL state was realized
notably in the 2D triangular system $\kappa$-(BEDT-TTF)$_{2}$Cu$_{2}$(CN)$_{3}$\cite{7-Kappa CuCn},
the Kagomé system ZnCu$_{3}$(OH)$_{6}$Cl$_{2}$ \cite{8-Kagome neutron 50mK,9-Kagome musr 50mK,10 Kagome NMR =000026 50mK,11 kagome O =000026  Cl NMR,12-M Vries},
and the hyperkagomé system Na$_{4}$Ir$_{3}$O$_{8}$ \cite{13 Na4Ir3O8 Takagi,14-Yogesh singh}.
A frustrated geometry and a low value of spin ($S=1/2$), which enhances
quantum fluctuations, help in stabilizing a QSL state \cite{1 Balents}.
However, even in the 6H-B and 3C phases of Ba$_{3}$NiSb$_{2}$O$_{9}$
(Ni$^{2+}$, $S=1$) \cite{15-Ba3NiSb2O9}, which have 2D edge-shared
triangular and \textcolor{black}{3D edge-shared tetrahedral lattices,}
respectively, a QSL state has been suggested.\textcolor{black}{{} Also,
the realization of a QSL state for Ba$_{3}$IrTi$_{2}$O$_{9}$ \cite{16-  Ba3Ir},
containing a diluted triangular lattice, and Ba$_{3}$YIr$_{2}$O$_{9}$}
(high pressure cubic phase) \cite{17- Ba3YIr2O9}, possibly suggests
the importance of further neighbor interaction and/or deviations from
the Heisenberg model. \textcolor{black}{Recent experimental/theoretical
results suggest that disorder might even drive the QSL state}\textcolor{red}{{}
}\textcolor{black}{\cite{18- AFM melt,19-Mila}.} The unconventional
nature of its elementary excitations, which result from a \textcolor{black}{chargeless
sector of spin-1/2 fermions (commonly known as spinons)}\textit{\textcolor{black}{,
}}\textcolor{black}{also drew interest from theorists and experimentalists
\cite{1 Balents,20- Wen order,21-Lee spinon}.} \textcolor{black}{Spinons
are believed to form a Fermi surface in such Mott insulators and a
non-vanishing Sommerfeld coefficient ($\gamma$) appears to be a generic
feature for most of the previously discussed QSL} \cite{7-Kappa CuCn,13 Na4Ir3O8 Takagi,15-Ba3NiSb2O9,17- Ba3YIr2O9}. 

\begin{figure}[t]
\begin{centering}
\includegraphics[bb=10mm 0bp 1280bp 607bp,scale=0.35]{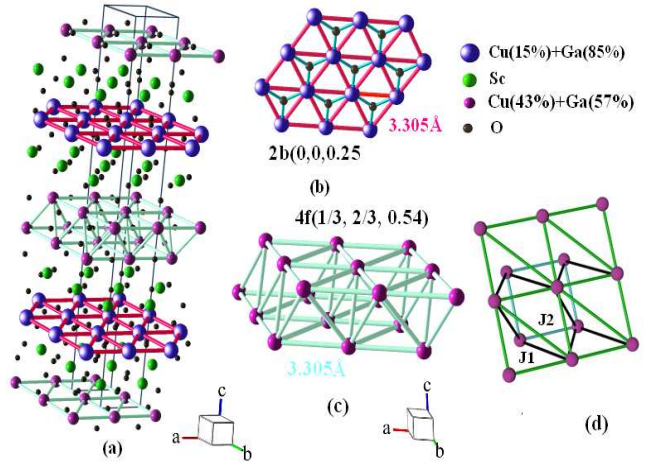}
\par\end{centering}

\protect\caption{(Color online)\label{Unit cell} (a) The unit cell of SGCO is shown.
(b) Edge-shared triangular plane of the 2b (0, 0, 1/4) sites is shown.
\textcolor{black}{(c) Triangular bi-planes from the 4f site (1/3,
2/3, 0.54) are shown. (d) an alternative representation of (c): to
maintain clarity, not all atoms are shown). The 2b and 4f sites are
shared by Cu and Ga (see legend and text). }}
\end{figure}

We have been exploring a variety of spin systems with the objective
of finding new QSL especially investigating triangular lattices. Herein,
we introduce a Cu$^{2+}(S=1/2)$ based potential QSL system Sc$_{2}$Ga$_{2}$CuO$_{7}$
(SGCO). This system was first reported by Kimizuka \textit{et al.
}\cite{22-SGCO} and the structure type was identified as similar
to Yb$_{2}$Fe$_{3}$O$_{7}$ \cite{23-Yb2Fe3O7}. However, no other
data have been reported up to now. Here we report a thorough investigation
of the structure and magnetic properties of SGCO by\textcolor{black}{{}
synchrotron x-ray diffraction (xrd), neutron diffraction (ND), susceptibility,
and specific heat measurements. }In the Yb$_{2}$Fe$_{3}$O$_{7}$
structure as applied to SGCO, the 4f sites (forming triangular bi-layers)
are expected to be occupied by Ga and 2b sites (forming single triangular
layers) by Cu (see \textcolor{black}{Fig. \ref{Unit cell}). However,
we found a large deviation from this expectation. The Ga and Cu occupancies
are not easy to obtain due to their similar scattering lengths for
neutron as well as similar atomic scattering factors for xrd. From
the xrd (synchrotron) data, where the x-ray energy was tuned to be
near the K-absorption edge of Cu, we could reliably estimate the 4f
and 2b site occupancies by Ga and Cu. Combined with our magnetization
and heat capacity data, this allows us to suggest that the magnetic
lattice of SGCO comprises of (i) triangular bi-planes, (double layers
of triangular bi-pyramids), of $S=1/2$ Cu which are nearly 50\% diluted
by Ga but give rise to spin liquid behavior and (ii) some uncorrelated
Cu located at a different site. For magnetic fields higher than 90
kOe the magnetic heat capacity exhibits a power law behavior with
an exponent close to 2 as in many QSL systems \cite{13 Na4Ir3O8 Takagi,15-Ba3NiSb2O9}. }

\textit{Sample preparation and experimental details}: \textcolor{black}{Several
batches of polycrystalline SGCO were synthesized by a conventional
solid-state reaction route (see Supplementary Material \cite{24-supplemental}).}
Xrd data were collected at 300 K using a PANalytical diffractometer
using Cu-K$\alpha$ radiation ($\lambda$ = 1.54182 $\textrm{\AA}$).
Synchrotron xrd data were measured at room temperature at the\textcolor{black}{{}
Materials Sciences Beamline \cite{25- Beamline}} of the Swiss Light
Source (SLS), with the wavelength 0.6204 Å, using the Mythen-II detector.\textcolor{red}{{}
}\textcolor{black}{Data were also taken at $\lambda=1.38455$ }Å (E
= 8.9548 keV) which is just below the K-absorption edge of Cu (E =
8.9789 keV).\textcolor{black}{{} Silicon powder was added to the substance
in order to substantially reduce the x-ray absorption in the sample.}\textcolor{red}{{}
}The ND measurements were carried out with the\textcolor{black}{{} HRPT
diffractometer \cite{26- HRPT DIFF}} at the SINQ neutron source ($\lambda$
= 1.49 Å) of Paul Scherrer Institut at 300 K and 1.5 K using a standard
orange cryostat.\textcolor{red}{{} }Magnetization $M$ measurements
were done using a Quantum Design SQUID VSM in the temperature $T$
range 1.8 - 400 K and in magnetic fields $H$ upto 70 kOe.\textcolor{black}{{}
The ac susceptibility measurements were performed on a dilution refrigerator
(0.3-7 K) and on a Quantum Design SQUID VSM in the $T$ range 2 -
30 K} \textcolor{black}{\cite{24-supplemental}}. The heat capacity
measurements were done in the $T$-range 0.35 - 295 K, using a Quantum
design PPMS.\textcolor{red}{{} }

\textit{\textcolor{black}{Structure and magnetic model}}\textcolor{black}{:
}The powder xrd and ND patterns of SGCO could be indexed within the
space group $P$$6$$_{3}/mmc$ (194) and correspond to the Yb$_{2}$Fe$_{3}$O$_{7}$
structure previously reported by Kimizuka \textit{et al}. \cite{22-SGCO,23-Yb2Fe3O7}.
\textcolor{black}{Our lab xrd data (not shown) did not show any impurity
peaks, however} in the much higher statistics synchrotron data we
found Sc$_{2}$O$_{3}$ ($\sim$ 1.2 wt.\%) and CuGa$_{2}$O$_{4}$
($\sim$ 0.5 wt.\%) impurities. A three phase Rietveld refinement
using FullProf \cite{27- FULL} was done on our synchrotron data to
obtain the lattice constants and the absolute amount of phases. The
refined synchrotron data with \textcolor{black}{($\lambda=0.62$}
\AA \textcolor{black}{)} and the extracted atomic positions for SGCO
are summarized in the Supplementary Material \cite{24-supplemental}.
The obtained lattice constants are $a=b=3.30395(3)\thinspace\textrm{Å}$
and $c=28.1116(3)\textrm{\thinspaceÅ}$, similar to those reported
by Kimizuka \textit{et al}. \cite{22-SGCO}. 

\begin{figure}
\begin{centering}
\includegraphics[scale=0.37]{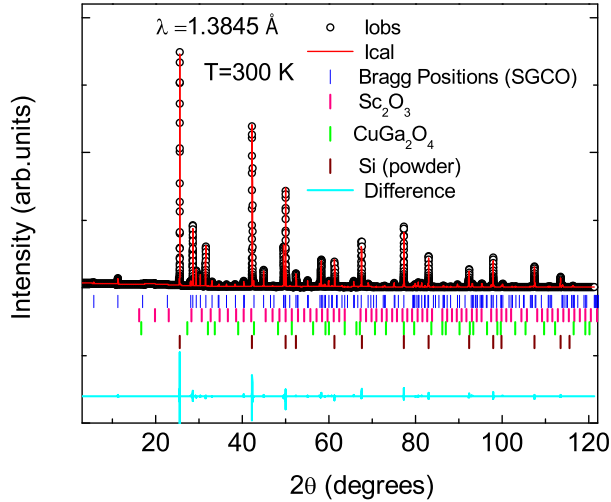}
\par\end{centering}

\protect\caption{\textcolor{black}{(Color online) \label{Resonant-xrays fig} Resonant
xrd (}SLS\textcolor{black}{) data (}$\lambda$\textcolor{black}{{} =
1.38455 $\textrm{\AA}$) for SGCO at 300 K refined with the space
group }$P$$6$$_{3}/mmc$\textcolor{black}{. Iobs and Ical represent
the experimental and Rietveld refined intensities, respectively. The
vertical blue, pink, green and brown bars depict the Bragg positions
for SGCO, Sc$_{2}$O$_{3}$, CuGa$_{2}$O$_{4}$ and Si (internal
standard), respectively, whereas flat cyan line shows Iobs- Ical.}}
\end{figure}

Since Cu and Ga have similar scattering lengths, both for x-rays and
for neutrons, possible site sharing between the two is not easily
addressed by analyzing the diffraction data.\textcolor{black}{{} For
instance, for x-rays ($\lambda=0.62$} \AA \textcolor{black}{) the
atomic scattering factors for Ga and Cu are different by only 7\%.
However, if one tunes the x-ray energy near the K-absorption edge
of Cu, the difference in scattering factors for Ga and Cu can be much
more. In our case ($\lambda=1.38455$} \AA \textcolor{black}{), the
scattering factors for Ga and Cu are different by 25\%. Xrd pattern
under these conditions is shown in Fig. \ref{Resonant-xrays fig}.
The obtained lattice constants are }\textit{\textcolor{black}{a}}\textcolor{black}{{}
= }\textit{\textcolor{black}{b}}\textcolor{black}{{} = 3.3045(2) $\textrm{\AA}$,
}\textit{\textcolor{black}{c}}\textcolor{black}{{} = 28.1129(1) $\textrm{\AA}$
(see }Supplementary Material \cite{24-supplemental}\textcolor{black}{{}
for details). We find that the 2b sites (0, 0, 1/4) contain 85\% non-magnetic
Ga$^{3+}$ and 15\% Cu$^{2+}$. The 4f sites (1/3, 2/3, $z$) contain
57\% Ga$^{3+}$and 43\% Cu$^{2+}$ ions. As shown in Fig. \ref{Unit cell}
(b) the edge-shared single-triangular planes at the 2b sites (0, 0,
1/4), along with the neighboring oxygens atoms make single layers
of corner-sharing triangular bi-pyramids (see} Supplementary Material
\cite{24-supplemental}\textcolor{black}{). On the other hand, the
4f sites (1/3, 2/3, 0.54) form triangular bi-planes (see Fig. \ref{Unit cell}
(c)) or rather double layers of triangular bi-pyramids when the oxygens
are included. In a bi-plane (4f sites), a triangular layer is shifted
with respect to the other such that the vertices of one layer are
at the centroids of alternate triangles of the other layer. This is
topologically equivalent to a honeycomb lattice with nearest ($J1$)
and next-nearest ($J2$) neighbor coupling and based on exchange paths
it is likely that ${\color{red}{\color{black}J}{\color{black}2>J1}}$,}\textcolor{red}{{}
}\textcolor{black}{see Fig. 1 (d).}

\textcolor{black}{The preference for Ga for the 2b sites is supported
by bond-valence sum (BVS) calculations. The BVS for Ga$^{3+}$ is
2.91 on 2b sites and 2.69 on 4f sites, making 2b the preferred site
for Ga$^{3+}$. Likewise the BVS for Cu is 2.54 on 2b sites and 2.35
on 4f sites, giving a slight preference for Cu$^{2+}$ to occupy 4f
sites, in agreement with the results from the structural refinements. }

\begin{figure}
\begin{centering}
\includegraphics[clip,scale=0.36]{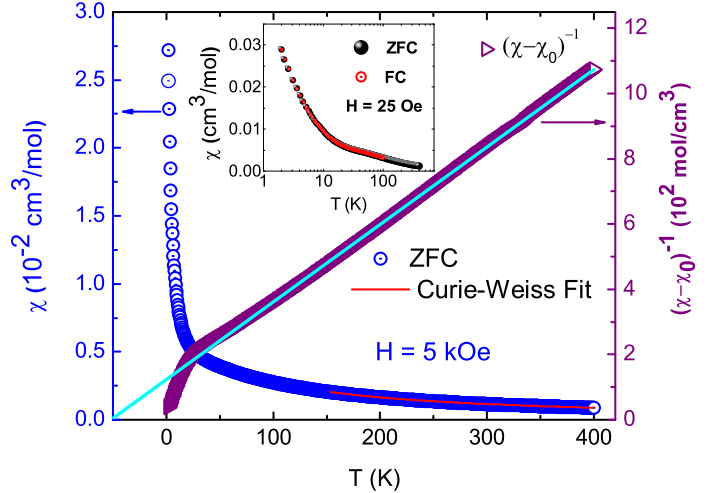}
\par\end{centering}

\protect\caption{(Color online) \label{Susceptibility}The left $y$-axis shows the
$T$-dependence of $\chi$ (blue open circles) measured in $H$ =
5 kOe and the right $y$-axis shows the inverse $\chi$ plot (purple
open triangles) free from the $T$-independent $\chi$. The Curie-Weiss
fit is shown in the $T$-range 150 K - 400 K with a red solid line
and the inset depicts the absence of any ZFC/FC bifurcation in $\chi(T)$
in $H$ = 25 Oe. }
\end{figure}

The dc susceptibility ($\chi=\frac{M}{H}$) of SGCO varies in a Curie-Weiss
manner $\chi=\chi_{0}+C/(T-\theta_{CW})$ down to 100 K, \textcolor{black}{where
$\chi_{0}$ (-4.451$\times$10$^{-5}$ cm$^{3}$/mol), $C$ = 0.41
cm$^{3}$ K/mol, and $\theta_{CW}$ $=$ -44 K} denote the $T$-independent
$\chi$, the Curie constant, and the Curie-Weiss temperature, respectively.
The negative \textcolor{black}{$\theta_{CW}$} suggests antiferromagnetic
correlations and the inferred effective moment $\mu_{eff}\approx\sqrt{8C}=1.79\pm$
0.09 $\mu_{B}$ is close to that for an $S=1/2$ moment. Measurements
performed in a low field (25 Oe) and down to 2 K do not exhibit any
anomaly or bifurcation between zero field cooled (ZFC) and field cooled
(FC) data (inset of Fig. \ref{Susceptibility}). \textcolor{black}{AC
$\chi$ measurements in the $T$-range 0.3-30 K do not show any anomaly
or frequency dependence (see the Supplemental material \cite{24-supplemental}).}
\textcolor{black}{This suggests the absence of LRO or any spin freezing.
}It is worth mentioning that the ND measurements carried out at 1.5
K did not show any signature of magnetic Bragg peaks, setting an upper
limit of 0.5 $\mu_{B}$ for the ordered moment in case of LRO. Further,
$^{71}$Ga NMR shift data \cite{28-khuntia_nmr} on SGCO indicate
a leveling off of $\chi$ below about 50 K and the absence of LRO.
While further evidence is needed to conclusively establish QSL behavior
in SGCO, the essential features of $\chi(T)$ in SGCO are as in other
QSL materials. 

\textcolor{black}{To estimate the fraction of paramagnetic spins in
SGCO, we have made use of the data from magnetic isotherms $M(H)$
at low-$T$ (1.8 - 4.5 K) since the Cu at the 2b sites is expected
to show paramagnetic Curie-like behavior while the contribution from
the possibly correlated triangular bi-planes might be much lower.
Our analysis (see the Supplemental material }\cite{24-supplemental}\textcolor{black}{)
of the $M(H)$ data is consistent with about 12\% of paramagnetic
Cu$^{2+}$($S=1/2$) spins which is not far from the value of 15\%
Cu at the 2b sites obtained from resonant xrd data. In In$_{2}$Ga$_{2}$CuO$_{7}$,
structurally similar to SGCO, Taetz }\textit{\textcolor{black}{et
al. }}\textcolor{black}{\cite{29-Teatz thesis} obtained approximately
10\% of paramagnetic $S=1/2$ impurities from a similar analysis.
Given the site occupancies obtained from resonant xrd results on SGCO,
the magnetic lattice (bi-layers from the atoms at 4f positions (1/3,
2/3, 0.54)) is more than 50\% diluted. While this is beyond the percolation
threshold of a 2D triangular lattice, for the bi-layer configuration
(or the honeycomb lattice as mentioned before) here, the magnetic
connectivity might still be maintained. }

\textcolor{black}{Based on our experiments, we suggest that SGCO appears
to consist of (i) 10-15\% of the 2b sites having $S=1/2$ paramagnetic
moments and (ii) a nearly equal mixture of $S=1/2$ Cu$^{2+}$ and
$S=0$ Ga$^{3+}$ at the 4f sites which magnetically form the triangular
bi-planes. While it could be that the site occupation is correlated,
maintaining the magnetic connectivity in the triangular bi-layers,
it is surprising that there is no spin-freezing even after large dilution.
One may speculate that the proximity of the site occupation to a percolation
threshold may promote long range spin singlets as in an algebraic
spin liquid \cite{1 Balents}.}

\begin{figure}[h]
\begin{centering}
\includegraphics[scale=0.4]{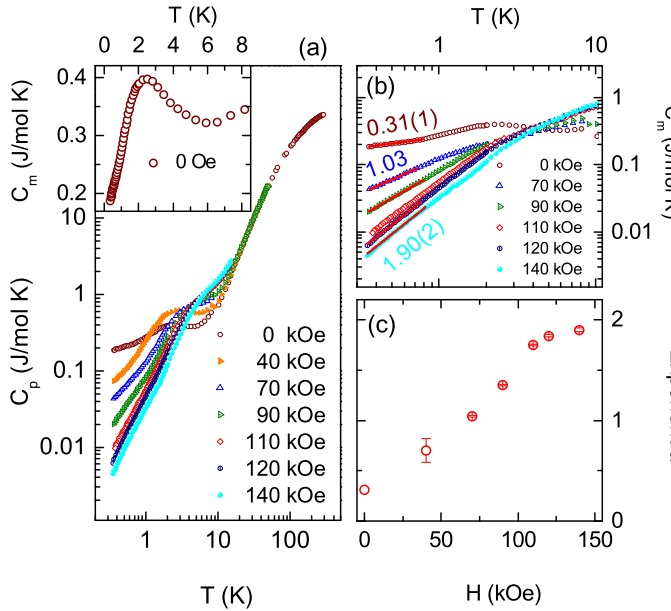}
\par\end{centering}

\protect\caption{(Color online) \label{Cp (T)} (a) The $C$$_{p}(T)$ data (log-log
scale) for various $H$ (up to 140 kOe) are shown. Inset depicts the
appearance of a broad maximum in $C_{m}$, in $H$ = 0. (b) Power
law fits (see text) of $C_{m}$ are shown as solid lines. (c) Variation
of exponent $\alpha$ with $H$ is shown. Note that, for $H$ > 40
kOe the size of the error bars is equal to the size of the symbols.}
\end{figure}

\textit{\textcolor{black}{Heat capacity}}\textcolor{black}{:} To obtain
further insight into the low-energy excitations of SGCO, we measured
its specific heat $C_{p}(T)$ (Fig. \ref{Cp (T)}) in the $T$-range
0.35 - 300 K under various $H$ (0 kOe - 140 kOe). We found no signature
of LRO down to 0.35 K. However, below about 15 K the $C$$_{p}(T)$
displays a dependence on $H$ which is suggestive of a Schottky anomaly.
\textcolor{black}{This Schottky contribution $C$$_{Schottky}$ most
likely arises from the paramagnetic Cu$^{2+}$ spins at the 2b sites.
Note that our $M(H)$ data analysis revealed a $\sim$12\% contribution
from paramagnetic $S=1/2$ species. }We then analyzed $C_{p}(T)$
as a combination of \textcolor{black}{$C$$_{Schottky}$} and \textcolor{black}{$C$$_{lattice}$}
(the lattice heat capacity) in addition to a magnetic heat capacity
$C_{m}$ (see \textcolor{black}{Supplemental Material} \cite{24-supplemental}).
The $C_{m}$ (which, we believe, comes from the correlated spins of
the triangular bi-planes) is obtained after subtracting \textcolor{black}{$C$$_{lattice}$}
and \textcolor{black}{$C$$_{Schottky}$} and is shown in Fig. \ref{Cp (T)}
(b). 

\textcolor{black}{{} As shown in the inset of Fig. \ref{Cp (T)} (a),
$C_{m}$($T$) shows a broad maximum (perhaps not related to any phase
transition) at around 2.5 K in $H$ = 0. The appearance of a broad
maximum is common for highly frustrated spin systems and has previously
been noticed for NiGa$_{2}$S$_{4}$ \cite{30- NiGa2S4}, Na$_{4}$Ir$_{3}$O$_{8}$
\cite{13 Na4Ir3O8 Takagi}, Ba$_{3}$CuSb$_{2}$O$_{9}$ \cite{31-Ba3CuSb2O9},
and Ba$_{3}$NiSb$_{2}$O$_{9}$ \cite{15-Ba3NiSb2O9} However, its
position in $T$ (in comparison to the strength of the exchange coupling)
varies from material to material. }

Finally the low-$T$ $C_{m}$ data ($T$ < 1 K) were fitted to a power-law
$C_{m}=\gamma T^{\alpha}$(where $\gamma$ is a constant), to infer
about the magnetic excitations (see Fig. \ref{Cp (T)} (b)). We found
that in SGCO, $\alpha$ increases from 0.3 to 1.9 with increasing
$H$.\textcolor{green}{{} }\textcolor{black}{The error bars for $\alpha$
in Fig. \ref{Cp (T)} (c) were arrived at by assuming a $\pm50$\%
uncertainty in $C$$_{Schottky}$. The value of $\alpha$ is robust
and reliable for $H>40$ kOe in the $T$-range 0.35-1 K, and attains
a value 1.9(2) for $H>90$ kOe.} An exponent close to 2 might be an
indication of a QSL and has previously been observed for the hyperkagom\textcolor{black}{é}
compound Na$_{4}$Ir$_{3}$O$_{8}$ \cite{13 Na4Ir3O8 Takagi} as
also the 3C phase of Ba$_{3}$NiSb$_{2}$O$_{9}$ \cite{15-Ba3NiSb2O9}.
\textcolor{black}{In the case of a Kagomé lattice, a theoretical approach
(spin-1/2 spinons obeying the Dirac spectrum \cite{32- Dirac}) leads
to a $T$$^{2}$ dependence of specific heat. However, there could
be other explanations.} \textcolor{black}{Recently, randomness-induced
QSL (also called random singlet) has been suggested for organic triangular
salts \cite{33-random singlet 1} as also for inorganic ones such
as herbertsmithite \cite{34-random singlet 2} with the Kagomé structure.
In herbertsmithite, a Zn/Cu disorder might cause a random Jahn-Teller
distortion of the {[}Cu(OH)$_{6}${]}$^{4-}$ giving rise to a random
modification of the exchange coupling in the Kagomé layer. For the
resulting gapless QSL state, a $T$-linear }$C_{m}$\textcolor{black}{{}
is expected. In the SGCO case, a similar Ga/Cu disorder exists which
might drive the QSL state. However, In SGCO (in zero field), the power
law exponent of }$C_{m}$\textcolor{black}{{} is 0.3. This amounts to
a divergence of }$C_{m}/T$\textcolor{black}{{} at low-$T$. Such a
divergence in Pr$_{2}$Ir$_{2}$O$_{7}$ (frustrated moments in a
metallic background; different than SGCO) has been seen as an indication
of proximity to a quantum critical point QCP \cite{35-pr2ir2o7}.
For SGCO, with the increasing $C_{m}$($T$) exponent (from 0.3 to
1.9) with field, one might speculate that the system moves from a
QCP to deeper into the insulating QSL region with field. A cautionary
note is that the inferred low-field }$C_{m}$\textcolor{black}{{} might
be affected by possible interaction between the correlated spins and
the orphan spins. }

\textcolor{black}{It is worth mentioning} that in our $C_{m}$ data
at $H=140$ kOe, $\gamma$$\sim34$ mJ/mol K$^{2.9}$. A non zero
value of $\gamma$, in general, indicates the presence of low-energy
(likely) gapless excitations for these Mott insulators at low-$T$.
While t\textcolor{black}{he reason for the field-induced suppression
of the $C_{m}$ (suggesting a suppression of magnetic excitations)
is not clear at this point, it could arise from a freezing out of
any interaction between the orphan spins (Cu$^{2+}$ in the 2b sites)
and the triangular bi-layers.} The estimated entropy change $\Delta S$
(see Supplemental Material \cite{24-supplemental}) by integrating
the $C_{m}/T$ versus $T$ data (\textcolor{black}{in zero field})
yields 1.2 J/mol K, which is nearly 20\% of that expected for a $S=1/2$
moment. This value will be even smaller when the data at the highest
field are considered. The significantly smaller $\Delta S$ is suggestive
of a large residual entropy at low temperatures, further suggesting
a QSL ground state in SGCO.\textcolor{red}{{} }

\textit{Summary}: Using the results obtained from various experimental
probes such as xrd, ND, dc/ac $\chi(T)$, $M(H)$, and $C_{p}(T)$,
we have explored the properties of SGCO\textcolor{black}{. Our resonant
xrd measurements carried out at the K-absorption edge of Cu allow
us to infer the site occupancies of Cu and Ga at the 2b sites (Cu:Ga
= 0.15:0.85) and the 4f sites (Cu:Ga = 0.43:0.57). We suggest that
the magnetic lattice can be viewed as a combination of (i) highly
depleted triangular bi-layers (or honeycomb layers with nearest as
also next-nearest neighbor interactions) giving rise to correlated
behavior and (ii) about 15\% Cu spins which are paramagnetic.} Our
ND data down to 1.5 K do not show any signature of LRO/spin freezing
and are consistent with the dc/ac $\chi(T)$ measurements. \textcolor{black}{Further,
the $C_{p}(T)$ data indicate absence of any transition down to 0.35
K. They are also indicative of the presence of low-energy excitations
and for high fields yield $C$$_{m}$$\propto$ $T$$^{2}$ below
1 K. Taken together, the apparent lack of LRO or spin freezing and
the existence of low-energy (likely gapless) excitations could be
indicative of a QSL ground state. Often QSL candidates with site-disorder
have been found to display spin freezing. It is therefore unusual
to find the combined lack of order and spin-freezing in a system so
heavily diluted as the triangular bi-layers are in this compound.
This could be a result of the dilution approaching a percolation threshold
and one may speculate if the unusual QSL behavior could arise close
to such thresholds. There is also recent work \cite{19-Mila} which
proposes disorder driven spin-orbital liquid in such systems.} Another
possibility is disorder induced bond randomness which can give rise
to a random singlet QSL as has been suggested in herbertsmithite \cite{34-random singlet 2}.
Further low-$T$ magnetization and local probe investigations such
as $\mu$SR would be useful to explore the properties of SGCO in greater
details.

\textit{Acknowledgements:} \textcolor{black}{We thank Department of
Science and Technology, Govt. of India and the Indo-Swiss joint research
programme, the Swiss National Science Foundation and its SINERGIA
network MPBH for financial support.}\textcolor{blue}{{} }R. Kumar acknowledges
CSIR, India for awarding him a research fellowship and B. Koteswararao
thanks DST INSPIRE fellowship to carry out the research work. This
work is partly based on experiments performed at the Swiss spallation
neutron source SINQ, Paul Scherrer Institute, Villigen, Switzerland.

\paragraph{...........................................................................}
\begin{lyxlist}{00.00.0000}
\item [{\textbf{\textcolor{red}{Supplemental}}}] \textbf{\textcolor{red}{material
for }}\textbf{\textcolor{red}{\uline{Sc$_{2}$Ga$_{2}$CuO$_{7}$:
A possible Quantum spin liquid near the percolation threshold}}}
\end{lyxlist}

\subsection{Sample preparation}

\textcolor{black}{Stoichiometric mixtures of pre-dried CuO (99.995\%
purity), Ga$_{2}$O$_{3}$(99.99\% purity), and Sc$_{2}$O$_{3}$(99.9\%
purity) were ground homogeneously in an agate mortar and then pelletized.
The pellet was placed in a high density alumina crucible and fired
at 950$^{\circ}$C for 24 hours in a Carbolite tubular furnace. The
pellet was further reground, pelletized, and then fired two times
at 1250$^{\circ}$C for 24 hours each. Several such batches were prepared
in this study. The final product was light green in color.}

\subsection{X-ray (synchrotron) diffraction and neutron diffraction analysis }

\begin{figure}
\begin{centering}
\includegraphics[scale=0.35]{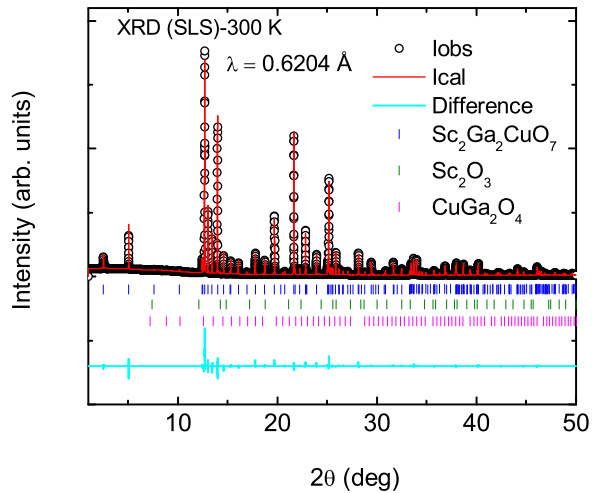}
\par\end{centering}

\protect\caption{\textcolor{black}{(Color online) \label{SLS-300K xray} X-ray (}synchrotron\textcolor{black}{)
diffraction data for SGCO recorded at 300 K with wavelength 0.6204
$\textrm{\AA}$. Iobs and Ical represent the experimental and Rietveld
refined intensities, respectively. The vertical blue, pink and green
bars depict the Bragg positions for Sc$_{2}$Ga$_{2}$CuO$_{7}$,
Sc$_{2}$O$_{3}$ and CuGa$_{2}$O$_{4}$, respectively, whereas flat
cyan line shows the intensity difference Iobs- Ical.}}
\end{figure}

\begin{figure}
\begin{centering}
\includegraphics[scale=0.5]{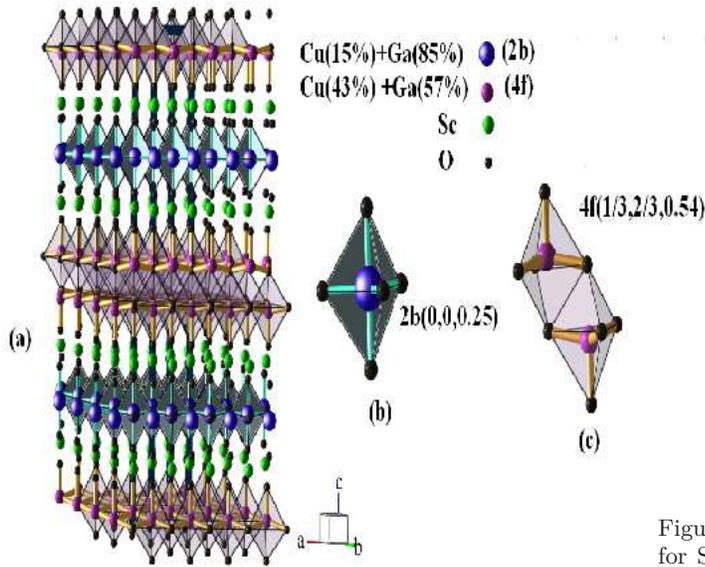}
\par\end{centering}

\protect\caption{(Color online). \label{SGCO STRUCTURE} (a) Local arrangement of atoms
in a unit cell of SGCO is shown. Blue and magenta color balls each
correspond to a fractional occupation by the Cu and Ga atoms as mentioned
in the legend. Figures 1 (b) and 1 (c) represent the Cu/Ga environments
at the 2b (\textcolor{black}{corner-sharing triangular bipyramids)
and the} 4f (\textcolor{black}{double layers of triangular bipyramids})
sites, respectively. }
\end{figure}

Figure \ref{SLS-300K xray} depicts the x-ray diffraction pattern
obtained for SGCO using synchrotron data with $\lambda=0.6204$ $\textrm{\AA}$.\textcolor{black}{{}
In the ideal case (no site sharing), the Cu and Ga atoms are expected
to form single and double triangular lattices comprising of atoms
at 2b (0, 0, 1/4) and 4f (1/3, 2/3, z) sites, respectively. However,
because of the similar ionic radii of Cu$^{2+}$ and Ga$^{3+}$ ions,
a distribution of both might be expected at the 2b (0, 0, 1/4) and
the 4f (1/3, 2/3, z) sites. The atomic coordinates refined from the
x-ray (synchrotron: SLS) diffraction data measured with }$\lambda$
= 0.6204 $\textrm{\AA}$\textcolor{black}{{} at 300 K for SGCO are given
in Table \ref{SLS 300K tab}. Since Cu (Z = 29) and Ga (Z = 31) have
nearly the same atomic scattering factors it is difficult to estimate
with high confidence, their distribution with the present x-ray (synchrotron:
}$\lambda$ = 0.6204 $\textrm{\AA}$\textcolor{black}{) analysis.
If the Ga:Cu ratio were left as a free parameter, the best refinement
of the synchrotron (SLS) data was obtained with the site 2b (0, 0,
1/4) mostly occupied by Ga$^{3+}$ and the 4f (1/3, 2/3, z) site having
both Cu$^{2+}$ and Ga$^{3+}$ ions, which is quite opposite to the
isostructural compound Yb$_{2}$Fe$_{3}$O$_{7}$ where the divalent
Fe$^{2+}$ and the trivalent Fe$^{3+}$ ions sit at 2b (0, 0, 1/4)
and 4f (1/3, 2/3, z) sites, respectively. The refinement quality factors
described by $R_{p}$, $R_{exp}$, $R{}_{wp}$ and $\chi^{2}$ are
obtained to be $3.56\%$, $0.34\%$, $3.50\%$ and $108$, respectively.
The local environments of Cu$^{2+}$ ions at 2b/4f site can be best
viewed as single layers of corner-sharing triangular bi-pyramids/double
layers of triangular bi-pyramids, as seen in Fig. \ref{SGCO STRUCTURE}.
As far as the magnetic lattice is concerned, the latter amounts to
a (heavily depleted) honeycomb lattice with the presence of both nearest
and next-nearest neighbor interactions.}

\textcolor{black}{We then collected x-ray diffraction data near the
K absorption edge of Cu to get a more reliable distribution of Cu
and Ga atoms at 2b and 4f sites, as the contrast between the atomic
scattering factors of Ga and Cu is higher at this wavelength. From
a refinement of our resonant x-ray data with }$\lambda=1.38455$ $\textrm{\AA}$\textcolor{black}{{}
(refined positions and occupancies are in Table \ref{Atomic pos}),
we obtain that the 2b sites (0, 0, 1/4) are mostly occupied with non-magnetic
Ga$^{3+}$ ions (85\% Ga$^{3+}$ and 15\% Cu$^{2+}$), whereas the
4f sites (1/3, 2/3, z), which form a network of double layers of triangular
bi-pyramids (or honeycomb layers as mentioned above), is populated
by Ga$^{3+}$ and Cu$^{2+}$ ions with 57\% Ga$^{3+}$ and 43\% Cu$^{2+}$,
see Fig. \ref{SGCO STRUCTURE}. The goodness of fit parameters R$_{p}$,
R$_{wp}$, R$_{exp}$ and $\chi^{2}$ are found to be 2.67\%, 2.94\%,
0.34\% and 75, respectively. }

\begin{table}
\protect\caption{\label{SLS 300K tab}\textcolor{blue}{{} }\textcolor{black}{Atomic positions
obtained after Rietveld refinement of the x-ray (synchrotron:SLS)
data at 300 K with }$\lambda=0.6204$$\textrm{ \AA}$\textcolor{black}{. }}

\centering{}%
\begin{tabular}{|c|c|c|c|c|c|}
\hline 
Atom & site & x & y & z & occ\tabularnewline
\hline 
\hline 
\textcolor{black}{Sc} & \textcolor{black}{4f} & \textcolor{black}{1/3} & \textcolor{black}{2/3} & \textcolor{black}{0.14877(2)} & \textcolor{black}{1}\tabularnewline
\hline 
\textcolor{black}{Cu} & \textcolor{black}{2b} & \textcolor{black}{0.0} & \textcolor{black}{0.0} & \textcolor{black}{1/4} & \textcolor{black}{0.252(2)}\tabularnewline
\hline 
\textcolor{black}{Ga} & \textcolor{black}{2b} & \textcolor{black}{0.0} & \textcolor{black}{0.0} & \textcolor{black}{1/4} & \textcolor{black}{0.744(2)}\tabularnewline
\hline 
\textcolor{black}{Ga} & \textcolor{black}{4f} & \textcolor{black}{1/3} & \textcolor{black}{2/3} & \textcolor{black}{0.54478(1)} & \textcolor{black}{0.312(3)}\tabularnewline
\hline 
\textcolor{black}{Cu} & \textcolor{black}{4f} & \textcolor{black}{1/3} & \textcolor{black}{2/3} & \textcolor{black}{0.54478(1)} & \textcolor{black}{0.69(3)}\tabularnewline
\hline 
\textcolor{black}{O1} & \textcolor{black}{4f} & \textcolor{black}{0.0000} & \textcolor{black}{0.0000} & \textcolor{black}{0.18007(6)} & \textcolor{black}{1}\tabularnewline
\hline 
\textcolor{black}{O2} & \textcolor{black}{4f} & \textcolor{black}{1/3} & \textcolor{black}{2/3} & \textcolor{black}{0.61130(6)} & \textcolor{black}{1}\tabularnewline
\hline 
\textcolor{black}{O3} & \textcolor{black}{4f} & \textcolor{black}{1/3} & \textcolor{black}{2/3} & \textcolor{black}{0.03325(7)} & \textcolor{black}{1}\tabularnewline
\hline 
\textcolor{black}{O4} & \textcolor{black}{2c} & \textcolor{black}{1/3} & \textcolor{black}{2/3} & \textcolor{black}{1/4} & \textcolor{black}{1}\tabularnewline
\hline 
\end{tabular}
\end{table}

\begin{table}
\protect\caption{\textcolor{black}{\label{Atomic pos}} Crystal structure parameters
of SGCO as refined from\textcolor{black}{{} synchrotron data ($\lambda=1.38455\textrm{ \AA}$)}
at 300 K with space group $P$$6$$_{3}/mmc$. The occupancies of
Cu and Ga at both sites have been constrained to ensure the ratio
of total amounts of Cu and Ga to be 1:2.}

\centering{}%
\begin{tabular}{|c|c|c|c|c|c|}
\hline 
Atom & site & $x$ & $y$ & $z$ & occupancy\tabularnewline
\hline 
\hline 
\textcolor{black}{Sc} & \textcolor{black}{4f} & \textcolor{black}{1/3} & \textcolor{black}{2/3} & \textcolor{black}{0.14865(2)} & \textcolor{black}{1}\tabularnewline
\hline 
\textcolor{black}{Cu1/Ga1} & \textcolor{black}{2b} & \textcolor{black}{0} & \textcolor{black}{0} & \textcolor{black}{1/4} & \textcolor{black}{0.15(1)/0.85(1)}\tabularnewline
\hline 
\textcolor{black}{Cu2/Ga2} & \textcolor{black}{4f} & \textcolor{black}{1/3} & \textcolor{black}{2/3} & \textcolor{black}{0.54480(1)} & \textcolor{black}{0.43(1)/0.57(1)}\tabularnewline
\hline 
\textcolor{black}{O1} & \textcolor{black}{4f} & \textcolor{black}{0} & \textcolor{black}{0} & \textcolor{black}{0.17934(7)} & \textcolor{black}{1}\tabularnewline
\hline 
\textcolor{black}{O2} & \textcolor{black}{4f} & \textcolor{black}{1/3} & \textcolor{black}{2/3} & \textcolor{black}{0.61118(7)} & \textcolor{black}{1}\tabularnewline
\hline 
\textcolor{black}{O3} & \textcolor{black}{4f} & \textcolor{black}{1/3} & \textcolor{black}{2/3} & \textcolor{black}{0.03371(7)} & \textcolor{black}{1}\tabularnewline
\hline 
\textcolor{black}{O4} & \textcolor{black}{2c} & \textcolor{black}{1/3} & \textcolor{black}{2/3} & \textcolor{black}{1/4} & \textcolor{black}{1}\tabularnewline
\hline 
\end{tabular}
\end{table}

\begin{figure}[H]
\begin{centering}
\includegraphics[scale=0.35]{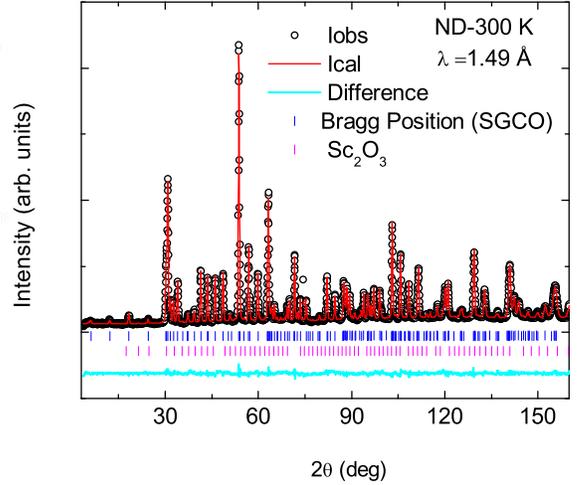}
\par\end{centering}

\protect\caption{(Color online) \label{ND-300K} Neutron diffraction data measured
for SGCO with $\lambda=1.49$ $\textrm{\AA}$ at 300 K are shown.
The open black circles and the red solid line represent the experimental
and refined data. The vertical blue and pink lines indicate the Bragg
positions for SGCO and Sc$_{2}$O$_{3}$, while flat cyan line shows
the difference plot between measured and refined intensities.}
\end{figure}

\begin{table}[H]
\protect\caption{\label{ND refined 300 =000026 1.5K} Refined positional parameters
and the occupancies for SGCO at 300 K under the space group $P$$6$$_{3}/mmc$
using the neutron diffraction data with $\lambda=1.49$ $\textrm{\AA}$. }

\centering{}%
\begin{tabular}{l>{\raggedright}p{1.5cm}>{\raggedright}p{1cm}>{\raggedright}p{1cm}>{\raggedright}p{3cm}l}
\hline 
 &  &  &  &  & \tabularnewline
Atom & Wyckoff

position & x  & y & z & Occupancy\tabularnewline
\hline 
Sc & 4f & 1/3 & 2/3 & 0.14861(5) & \textcolor{black}{0.93}\tabularnewline
Cu & 2b & 0 & 0 & 1/4 & \textcolor{black}{0.66}\tabularnewline
Ga & 2b & 0 & 0 & 1/4 & \textcolor{black}{0.34 }\tabularnewline
Cu & 4f & 1/3 & 2/3 & 0.54447(5) & \textcolor{black}{0.16 }\tabularnewline
Ga & 4f & 1/3 & 2/3 & 0.54447(5) & \textcolor{black}{0.84}\tabularnewline
O1 & 4f & 0 & 0 & 0.18087(7) & \textcolor{black}{1}\tabularnewline
O2 & 4f & 1/3 & 2/3 & 0.61201(6) & \textcolor{black}{1}\tabularnewline
O3 & 4f & 1/3 & 2/3 & 0.03320(10) & \textcolor{black}{1}\tabularnewline
O4 & 2c & 1/3 & 2/3 & 1/4 & \textcolor{black}{1}\tabularnewline
\hline 
\end{tabular}
\end{table}

\begin{figure}[H]
\begin{centering}
\includegraphics[scale=0.35]{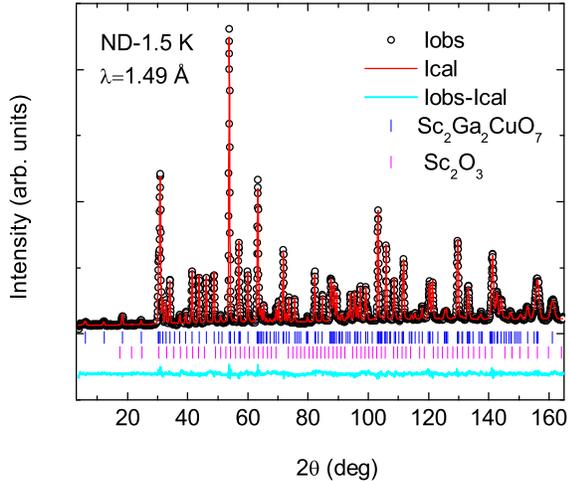}
\par\end{centering}

\protect\caption{(Color online) \label{ND-1.5K} Neutron diffraction data for SGCO,
collected at 1.5 K, with its Rietveld refinement is shown. Vertical
blue and pink bars depict the Bragg positions for SGCO (main phase)
and Sc$_{2}$O$_{3}$ (impurity phase $\sim$1.2\%).}
\end{figure}

\textcolor{black}{Similarly, the neutron diffraction measurements
performed at 300 K, shown in Fig}\textcolor{blue}{.}\textcolor{black}{{}
\ref{ND-300K}, also could not reliably infer the amount of site sharing
between Cu$^{2+}$and Ga$^{3+}$ions, because of a poor contrast between
their scattering lengths, 7.718 and 7.288 fm, respectively \cite{1-scattering lengths}.
Consequently, based on our neutron diffraction data alone we could
not have made a definitive statement about the site occupancy by Ga
and Cu at the 2b and 4f sites. The obtained lattice constants from
neutron diffraction data (}$\lambda$ = 1.49 $\textrm{\AA}$\textcolor{black}{)
are }$a=b=3.30459\textrm{\ensuremath{(4)} }$$\textrm{\AA}$ and $c=28.1183(4)$\textcolor{black}{{}
}$\textrm{\AA}$\textcolor{black}{, in good agreement with those obtained
from x-ray (synchrotron) data. The residual refinement factors $R_{p}$,
$R_{exp}$, $R{}_{wp}$ and $\chi^{2}$ for neutron refined powder
diffraction pattern for 300K data yield 3.61\%, 3.00\%, 4.73\% and
2.49, respectively. The refined atomic coordinates for SGCO at 300
K are listed in the Table \ref{ND refined 300 =000026 1.5K}. As it
is evident from the refinement result that the neutron diffraction
suggests a distribution of Cu and Ga quite opposite to what was determined
from resonant x-ray data. The neutron diffraction data collected at
1.5 K (}Fig. \ref{ND-1.5K}\textcolor{black}{) do not show the appearance
of any new peak and hence rule out the presence of magnetic ordering.}

\begin{figure}[H]
\begin{centering}
\includegraphics[scale=0.35]{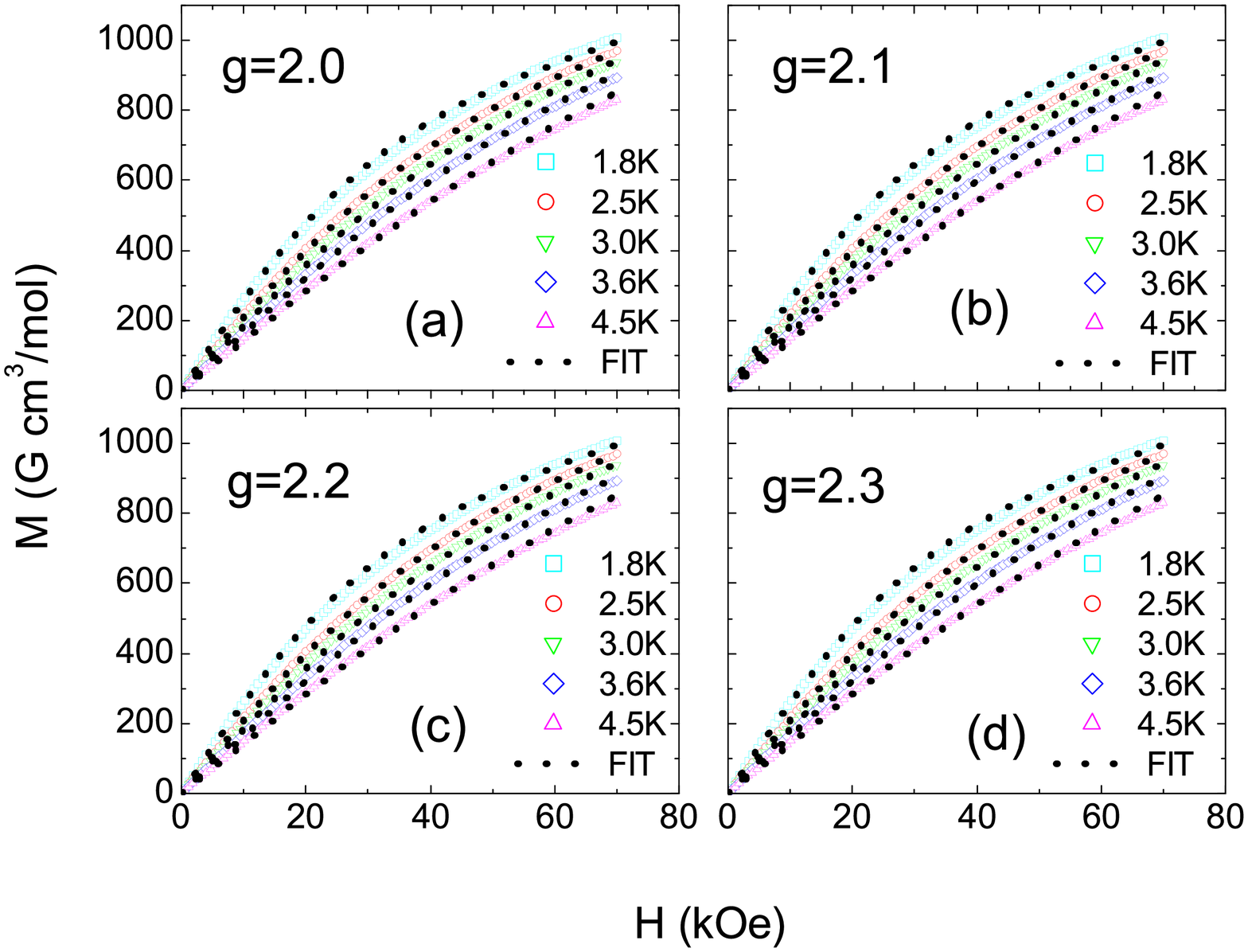}
\par\end{centering}

\protect\caption{(Color online) \label{Brill} The magnetization isotherms measured
in the field range (0-70 kOe) and in the temperature range 1.8-4.5
K are shown. The dotted broken lines are fits to an equation described
in the text. The panels (a), (b), (c), and (d) illustrate these fits
for various values of the spectroscopic splitting factor \textit{g}
= 2.0, 2.1, 2.2, and 2.3, respectively.}
\end{figure}

\subsection{Magnetization isotherm analysis}

\textcolor{black}{As it is evident from Fig. \ref{Brill}, below about
5 K magnetization isotherms start to show a curvature, which becomes
prominent at lower temperatures.} Measured magnetic isotherms in the
temperature range 1.8-4.5 K, where $\mu_{B}H$ is no longer much smaller
than $k_{B}T$, were analyzed within the framework of the following
equation: $M(H,T)=$ $\chi(T)H+N_{A}\mu_{B}f_{imp}S_{imp}g_{imp}B_{S}(g_{imp}S_{imp}H\mu_{B}/k_{B}(T-\theta_{imp}))$,
where $\chi$ represents the intrinsic susceptibility of the triangular
planes and is assumed constant in the 1.8-4.5 K interval, $f_{imp}$,
$g_{imp}$, $S_{imp}$, and $\theta_{imp}$ denote the impurity/disorder
concentration, the Lande \textit{g}-factor, spin, and correlation
temperature for impurity spins. The symbols $H$, $N_{A}$, $k_{B}$,
$\mu_{B}$, and $B_{S}$ refer to the magnetic field, the Avogadro
number, the Boltzmann constant, the Bohr magneton, and the Brillouin
function, respectively. To start with, there are five-parameters to
be varied and a simultaneous fitting of all these end up in nonphysical
values. So we first fixed $S$$_{imp}$ and \textit{g}$_{imp}$ to
be 1/2 and 2, respectively and thus reduced our variables from five
to three, \textit{i.e.}, $\chi$, $f$$_{imp}$ and $\theta_{imp}$.
To increase the reliability of our fitting parameters we simultaneously
fitted all magnetic isotherms, measured in the $T$-range 1.8 - 4.5
K. The outcomes of fittings for various assumed values of \textit{g}
are summarized in Table. \ref{Brill result}. As is evident from the
result, the value of $\theta_{imp}$ is nearly zero, which in turn
suggests the absence of magnetic correlations among impurity spins
so one can treat them as purely paramagnetic. The obtained free-spin
concentration is consistent with about 12\% antisite disorder.

\begin{flushright}
\begin{table}[H]
\begin{centering}
\protect\caption{\label{Brill result} Results of various fits to the magnetization
isotherm data with the equation described in text.}

\par\end{centering}

\centering{}%
\begin{tabular}{|>{\centering}p{1.5cm}|>{\centering}p{1.5cm}|r|>{\centering}p{2cm}|>{\centering}p{2cm}|}
\hline 
$S$$_{imp}$ & \textit{g}$_{imp}$ & $\chi$(cm$^{3}$/mol) & $\theta_{imp}$(K) & $f$$_{imp}$(\%)\tabularnewline
\hline 
\hline 
1/2 & 2.0 & 0.0051 & -0.22 & 11.60\tabularnewline
\hline 
1/2 & 2.1 & 0.0047 & -0.42 & 11.58\tabularnewline
\hline 
1/2 & 2.2 & 0.0043 & -0.63 & 11.59\tabularnewline
\hline 
1/2 & 2.3 & 0.0038 & -0.85 & 11.61\tabularnewline
\hline 
\end{tabular}
\end{table}

\par\end{flushright}

\subsection{A.C. susceptibility}

\begin{figure}[H]
\begin{centering}
\includegraphics[scale=0.35]{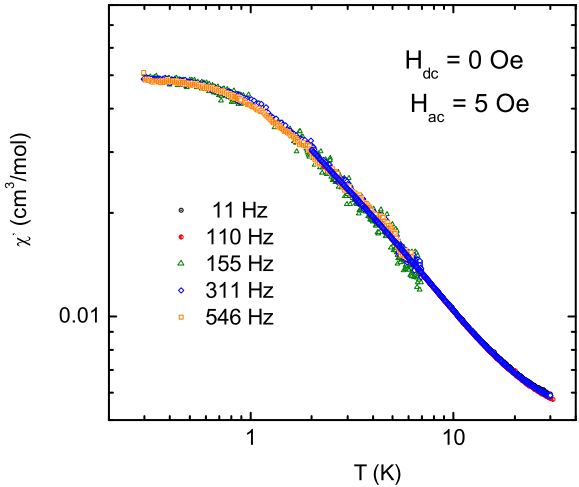}
\par\end{centering}

\protect\caption{\label{ac chi} Variation of the real part of the ac susceptibility
as a function of temperature (0.3-30 K) for several frequencies is
shown.}
\end{figure}

\textcolor{black}{AC susceptibility data were measured in the temperature
range 0.3 -7 K and 2- 30 K on a home-made susceptometer and a Quantum
Design SQUID VSM machine, respectively. The data were recorded keeping
the dc field zero and the ac pulse of amplitude 5 Oe. With the home-made
susceptometer we measured the data for three different frequencies
of 545.7 Hz, 311 Hz, and 155 Hz. In the temperature range 2-30 K on
Quantum design SQUID VSM machine we measured the data for five different
frequencies of 11 Hz, 110 Hz, 155 Hz, 311 Hz, and 546 Hz. The data
obtained with the home made susceptometer were multiplied by a scaling
factor (after subtracting a background) so as to match the data between
2 K and 7 K obtained using the Quantum Design SQUID VSM. The real
part of the susceptibility data is shown in Fig. \ref{ac chi}. These
data do not show any frequency dependence or any peak in the entire
measured temperature range and thus rule out the possibility of any
glassy behavior in this material within the precision of our measurements.}

\subsection{Magnetic specific heat analysis}

\begin{figure}[H]
\begin{centering}
\includegraphics[scale=0.35]{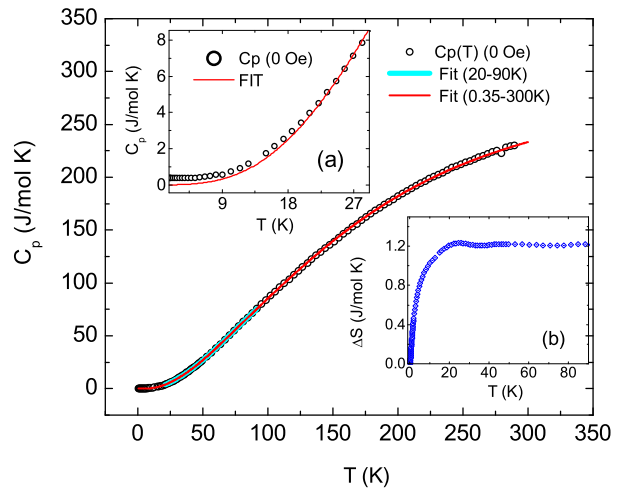}
\par\end{centering}

\protect\caption{(Color online) \label{Cm + Sm} Heat capacity at zero field is shown
with black open circles. Cyan and red solid lines represent initial
fit (20-90 K) to a combination of Debye and Einstein model and its
extrapolation in the entire temperature range (0.35-300 K), respectively.
Inset (a) highlights the low-temperature portion of the $C_{p}(T)$
data and inset (b) illustrates the entropy change at zero field.}
\end{figure}

\textcolor{black}{The $C$$_{p}(T)$ data, specific heat at constant
pressure, data obtained in zero magnetic field in the temperature
range 0.35-300 K are shown in Fig. \ref{Cm + Sm}. The total specific
heat of the system can be written as: $C$$_{p}(T)$ = $C$$_{lattice}$+
$C$$_{Schottky}$ + $C$$_{m}$. Here, $C$$_{lattice}$ is the lattice
specific heat and $C$$_{Schottky}$ originates from isolated paramagnetic
spins (Cu$^{2+}$ spins in this case) forming a two-level system.
We ascribe the remaining contribution $C$$_{m}$ to antiferromagnetically
interacting Cu$^{2+}$ spins residing in the 4f planes (bi-planes).
We now attempt to estimate $C$$_{lattice}$ and $C$$_{Schottky}$
so that the \textquotedbl{}intrinsic\textquotedbl{} contribution $C$$_{m}$
can be inferred.}

\textcolor{black}{To estimate the lattice contribution ($C$$_{lattice}$)
in the absence of a suitable non-magnetic analog for SGCO, we used
a combination of Debye and Einstein terms}, as written in Eqn. (1)
given below.

\begin{equation}
C_{lattice}(T)=C_{Debye}+C_{Einstein}
\end{equation}

where, $C_{Debye}=C_{D}\left[9k_{B}\left(\frac{T}{\theta_{D}}\right)^{3}\int\limits _{0}^{x_{D}}\frac{x^{4}e^{x}}{(e^{x}-1)^{2}}dx\right]$
and $C_{Einstein}=\sum C_{E_{i}}\left[3R\left(\frac{\theta_{E_{i}}}{T}\right)^{2}\frac{exp(\frac{\theta_{E_{i}}}{T})}{\left(exp(\frac{\theta_{E_{i}}}{T})-1\right)^{2}}\right]$

One formula unit of SGCO has twelve atoms and this in turn offers
one acoustic and eleven optical modes of atomic vibrations, in each
crystallographic direction. The first term (Debye integral) of the
Eqn. (1) takes care of the three acoustic modes of vibration whereas
the second term accounts for the contribution of thirty three optical
modes of phonons. This amounts to the conditions $C$$_{D}$= 1 and
$\sum$$C$$_{E_{i}}$=11 or $C$$_{D}$+ $\sum$$C$$_{E_{i}}$ =
12 for each direction. The data were found to fit well with weightage
factors corresponding to Debye and Einstein modes chosen in the ratio
1:1:4:6.\textcolor{black}{{} We, however, understand that this is just
a simple model to capture the basic characteristics of the complex
lattice dynamics. We initially fitted the $C$$_{p}(T)$ data to Eqn.
(1) in the temperature range 20-90 K and then extrapolated the curve
to cover the entire temperature range 0.35-300 K, as shown in the
inset (a) of Fig. \ref{Cm + Sm}. The fitting yields the Debye temperature
to be 159 K and Einstein temperatures to be 231 K, 308 K and 645 K,
respectively. Nevertheless, at very low temperatures ($T$$\ll\theta_{D}$:
Debye temperature) only the first term (acoustic branch of the phonon
spectrum) dominates as $T$$^{3}$ and Einstein weightage to the lattice
specific heat falls rapidly as exp(-$\theta_{D}/T$). }

\textcolor{black}{Next, we will estimate the Schottky contribution
$C$$_{Schottky}$ as follows. Based on magnetization isotherm analysis,
we modeled the Schottky contribution ($C$$_{Schottky}$) by fixing
the amount of free spins to 12\%. We used the two-level Schottky formula,
described by Eqn. 2 below, with $f=12$\% contribution from non-interacting
Cu-spins having a }\textit{\textcolor{black}{g}}\textcolor{black}{-value
of 2.} 

\begin{center}
\begin{equation}
C_{Schottky}=f\left[R(\frac{\triangle}{k_{B}T})^{2}(\frac{g_{0}}{g_{1}})\frac{exp(\frac{\triangle}{k_{B}T})}{[1+(\frac{g_{0}}{g_{1}})exp(\frac{\triangle}{k_{B}T})]^{2}}\right]
\end{equation}
 
\par\end{center}

Here, $\triangle$ is the energy separation under the influence of
an external magnetic field $H$, and $R$, $k{}_{B}$, $g{}_{0}$
and $g{}_{1}$ are the universal gas constant, the Boltzmann constant,
and the degeneracies of the two level system, respectively.\textcolor{red}{{} }

\begin{figure}
\begin{centering}
\includegraphics[scale=0.33]{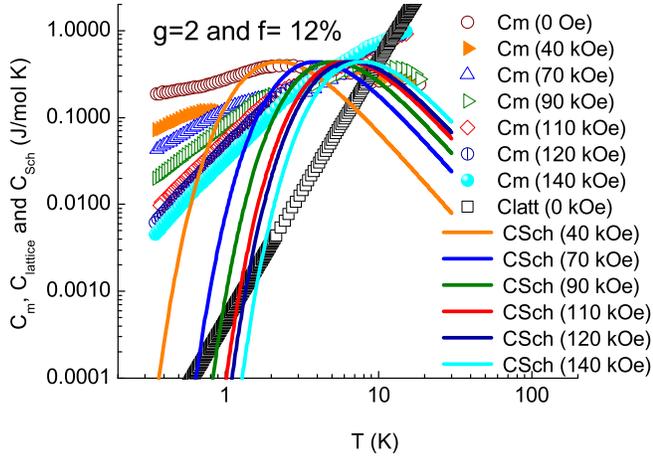}
\par\end{centering}

\protect\caption{(Color online)\label{fig:Cm+Sch+-Clatt} Individual contributions
of lattice \textcolor{black}{$C$$_{lattice}$} (open squares), Schottky
\textcolor{black}{$C$$_{Schottky}$} (various solid lines as per
field for f=12\%) and magnetic specific heats \textcolor{black}{$C$$_{m}$}
(other open and closed symbols) to the total specific heat are shown
in the field range 0-140 kOe.}
\end{figure}

\begin{figure}[H]
\begin{centering}
\includegraphics[scale=0.35]{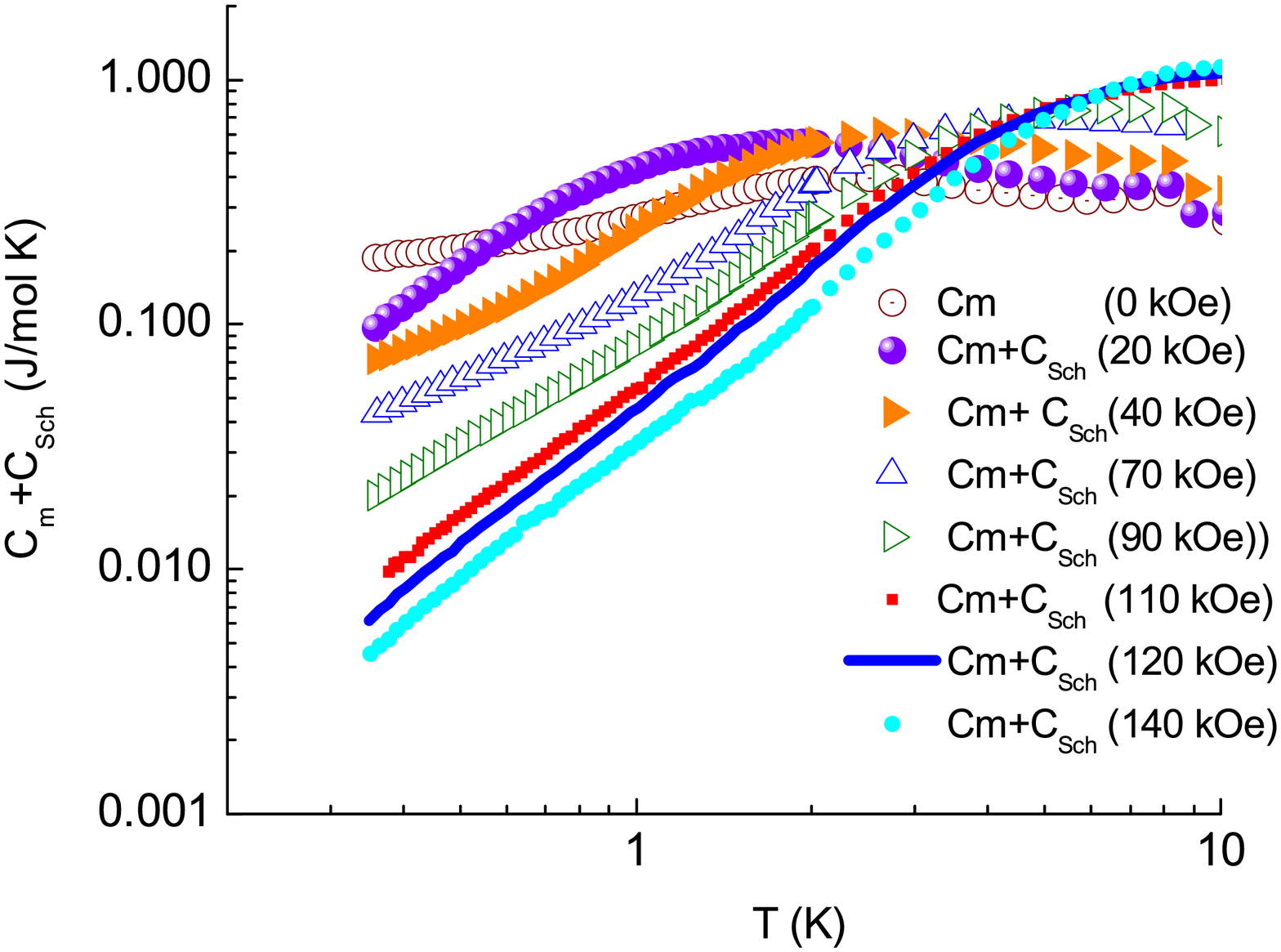}
\par\end{centering}

\protect\caption{(Color online)\label{Cm+Csch} A sum of magnetic specific heat \textcolor{black}{$C$$_{m}$}
and Schottky specific heat \textcolor{black}{$C$$_{Schottky}$} is
shown for fields 0 - 140 kOe as a function of temperature. }
\end{figure}

The \textquotedbl{}intrinsic\textquotedbl{} magnetic specific heat
\textcolor{black}{$C$$_{m}$ is thus obtained by subtracting $C$$_{lattice}$+
$C$$_{Schottky}$ from $C$$_{p}(T)$. Figure \ref{fig:Cm+Sch+-Clatt}
illustrates a plot of individual contributions of Schottky (for $g$
= 2 and $12$\% free spin contribution), lattice and magnetic specific
heats to the total specific heat data.} It is worth mentioning that
below about 1 K the lattice heat capacity as also the Schottky contribution
in fields beyond 90 kOe is utterly negligible in the total specific
heat and the determination of the magnetic specific heat (\textcolor{black}{$C$$_{m}$})
below 1 K is free from any uncertainty. Further, analyzing the magnetic
heat capacity data assuming a $\pm50$\% uncertainty in the Schottky
part does not change significantly the values of exponents for fields
$H$ > 40 kOe, and the exponents are found to be robust for $H$ >
40 kOe (see main text and Fig. 4 (c)). Note that Fig. \ref{Cm+Csch}
shows the lattice removed $C$$_{p}(T)$ data, a sum of both the magnetic
and Schottky specific heat, for SGCO at different magnetic fields. 

\begin{thebibliography}{10}
\bibitem{1 Balents} L. Balents, Nature \textbf{464}, 199 (2010).

\bibitem{2 Anderson} P. W. Anderson, Mater. Res. Bull. \textbf{8},
153 (1973).

\bibitem{3 NEEL Support} B. Bernu, P. Lecheminant, C. Lhuillier,
and L. Pierre, Phys. Rev. B \textbf{50} , 10048 (1994).

\bibitem{4 NEEL Support} R. R. P. Singh and D. A. Huse, Phys. Rev.
Lett. \textbf{68} , 1766 (1992 ).

\bibitem{5 NEEL Support} D. J. J. Farnell, R. F. Bishop, and K. A.
Gernoth, Phys. Rev. B \textbf{63} , 220402 R (2001). 

\bibitem{6 NEEL Support} L. Capriotti, A. E. Trumper, and S. Sorella,
Phys. Rev. Lett. \textbf{82} , 3899 (1999). 

\bibitem{7-Kappa CuCn} Y. Shimizu, K. Miyagawa, K. Kanoda, M. Maesato,
and G. Saito, Phys. Rev. Lett. \textbf{91}, 107001 (2003).

\bibitem{8-Kagome neutron 50mK} J. S. Helton, K. Matan, M. P. Shores,
E. A. Nytko, B. M. Bartlett, Y. Yoshida, Y. Takano, A. Suslov, Y.
Qiu, J.-H. Chung, D. G. Nocera, and Y. S. Lee, Phys. Rev. Lett. \textbf{98},
107204 (2007). 

\bibitem{9-Kagome musr 50mK} P. Mendels, F. Bert, M. A. de Vries,
A. Olariu, A. Harrison, F. Duc, J. C. Trombe, J. S. Lord, A. Amato,
C. Baines Phys. Rev. Lett. \textbf{98} , 077204 (2007).

\bibitem{10 Kagome NMR =000026 50mK} T. Imai, E. A. Nytko, B. M.
Bartlett, M. P. Shores, and D. G. Nocera, Phys. Rev. Lett. \textbf{100},
077203 (2008).

\bibitem{11 kagome O =000026  Cl NMR} A. Olariu, P. Mendels, F. Bert,
F. Duc, J. C. Trombe, M. A. de Vries, and A. Harrison, Phys. Rev.
Lett. \textbf{100}, 087202 (2008).

\bibitem{12-M Vries} M. A. de Vries, J. R. Stewart, P. P. Deen, J.
Piatek, G. N. Nilsen, H. M. Ronnow, and A. Harrison, Phys. Rev. Lett.
\textbf{103}, 237201 (2009).

\bibitem{13 Na4Ir3O8 Takagi} Y. Okamoto, M. Nohara, H. Aruga-Katori,
and H. Takagi, Phys. Rev. Lett. \textbf{99}, 137207 (2007). 

\bibitem{14-Yogesh singh} Y. Singh, Y. Tokiwa, J. Dong, and P. Gegenwart,
Phys. Rev. B \textbf{88, }220413(R) (2013).

\bibitem{15-Ba3NiSb2O9} J. G. Cheng, G. Li, L. Balicas, J. S. Zhou,
J. B. Goodenough, Cenke Xu, and H. D. Zhou, Phys. Rev. Lett. \textbf{107},
197204 (2011). 

\bibitem{16-  Ba3Ir} Tusharkanti Dey, A. V. Mahajan, P. Khuntia,
M. Baenitz, B. Koteswararao, and F. C. Chou, Phys. Rev. B \textbf{86},
140405(R) (2012).

\bibitem{17- Ba3YIr2O9} Tusharkanti Dey, A. V. Mahajan, R. Kumar,
B. Koteswararao, F. C. Chou, A. A. Omrani, and H. M. Ronnow, Phys.
Rev. B \textbf{88}, 134425 (2013). 

\bibitem{18- AFM melt} T. Furukawa, K. Miyagawa, T. Itou, M. Ito,
H. Taniguchi, M. Saito, S. Iguchi, T. Sasaki, and K. Kanoda, Phys.
Rev. Lett. \textbf{115}, 077001 (2015). 

\bibitem{19-Mila} A. Smerald and F. Mila, Phys. Rev. Lett. \textbf{115},
147202 (2015).

\bibitem{20- Wen order} X. G Wen, Phys. Rev. B\textbf{ 65}, 165113
(2002).

\bibitem{21-Lee spinon} T. K. Ng and P. A. Lee, Phys. Rev. Lett.
\textbf{99}, 156402 (2007).

\bibitem{22-SGCO} N. Kimizuka, T. Mohri, J. Solid State Chem. \textbf{60},
382 (1985). 

\bibitem{23-Yb2Fe3O7} N. Kimizuka, A. Takenaka, Y. Sasada, T. Katsura,
Solid State Comm.\textbf{ 15}, 1199 (1974).

\bibitem{24-supplemental} See supplemental material for sample preparation,
x-ray, neutron refinement, magnetic, and specific heat analysis.

\bibitem{25- Beamline} http://www.psi.ch/sls/ms/powder-diffraction

\bibitem{26- HRPT DIFF} \textcolor{black}{P. Fischer, G. Frey, M.
Koch, M. Könnecke, V. Pomjakushin, J. Schefer, R. Thut, N. Schlumpf,
R. Bürge, U. Greuter, S. Bondt, and E. Berruyer, Physica B }\textbf{\textcolor{black}{146}}\textcolor{black}{,
276 (2000).}

\bibitem{27- FULL} J. Rodríguez-Carvajal, Physica B. \textbf{192},
55 (1993).

\bibitem{28-khuntia_nmr}P. Khuntia \textit{et al}. unpublished.

\bibitem{29-Teatz thesis} T. Taetz\textcolor{black}{,} PhD thesis,
University of Cologne, Germany, (2008).

\bibitem{30- NiGa2S4} S. Nakatsuji, Y. Nambu, H. Tonomura, O. Sakai,
S. Jonas, C. Broholm, H. Tsunetsugu, Y. Qiu, and Y. Maeno , Science
\textbf{309}, 1697 (2005).

\bibitem{31-Ba3CuSb2O9} H. D. Zhou, E. S. Choi, G. Li, L. Balicas,
C. R. Wiebe, Y. Qiu, J. R. D. Copley, and J. S. Gardner\textcolor{black}{,}
Phys. Rev. Lett. \textbf{106}, 147204 (2011).

\bibitem{32- Dirac} Y. Ran, M. Hermele, P. A. Lee, and X.-G. Wen,
Phys. Rev. Lett. \textbf{98}, 117205 (2007).

\bibitem{33-random singlet 1} K. Watanabe, H. Kawamura, H. Nakano,
and T. Sakai, J. Phys. Soc. Japan \textbf{83}, 034714 (2014).

\bibitem{34-random singlet 2} T. Shimokawa, K. Watanabe, and H. Kawamura,
arXiv:1506.03576v2.

\bibitem{35-pr2ir2o7} Y. Tokiwa, J. J. Ishikawa, S. Nakatsuji, and
P. Gegenwart, Nature Materials \textbf{13}, 359 (2014). 

\end{thebibliography}

\begin{thebibliography}{10}
\bibitem{1-scattering lengths}  See for neutron scattering lengths:
http://www.ncnr.nist.gov/resources/n-lengths/\end{thebibliography}
\end{document}